\documentclass{jltp}

\usepackage{graphicx}

\title{Tuning correlation effects with electron-phonon interactions}
\author{J.P.Hague and N.d'Ambrumenil}
\address{Department of Physics, University of Warwick, CV4 7AL, U.K.}

\date{June 2005}

\begin{document}

\maketitle

\begin{abstract}
We investigate the effect of tuning the phonon energy on the
correlation effects in models of electron-phonon interactions using
DMFT. In the regime where itinerant electrons, instantaneous
electron-phonon driven correlations and static distortions compete on
similar energy scales, we find several interesting results including
(1) A crossover from band to Mott behavior in the spectral function,
leading to hybrid band/Mott features in the spectral function for
phonon frequencies slightly larger than the band width. (2) Since the
optical conductivity depends sensitively on the form of the spectral
function, we show that such a regime should be observable through the
low frequency form of the optical conductivity. (3) The resistivity
has a double kondo peak arrangement [Published as J. Low. Temp. Phys. {\bf 140} pp77-89 (2005)].

\end{abstract}

\section{Introduction}
\label{limitbehaviour}

Mounting experimental evidence from high-Tc cuprates
\cite{lanzara2001a}, nickelates \cite{tranquada2002a}, manganites
\cite{zhao1996a,millis1996b} and other interesting materials suggests
that large electron-phonon interactions may play a more important role
in the physics of strongly correlated electron systems than previously
thought. Migdal-Eliashberg and BCS theories have proved extremely
successful in describing the effects of phonons in many
materials. However, if the coupling between electrons and the
underlying lattice is large, and/or the phonons can not be treated
within an adiabatic approximation, conventional approaches fail.

The Holstein model contains most of the fundamental physics of the
electron-phonon problem \cite{holstein1959a}. Tight-binding electrons
are coupled to the lattice through a local interaction with Einstein
modes. For large phonon frequencies, electrons interact with a
strongly correlated Hubbard-like attraction, while for small phonon
frequencies the lattice gives rise to a static potential which is
essentially uncorrelated. Between these two extreme limits of
correlated and uncorrelated behavior, levels of correlation are tuned
by the size of the phonon frequency and novel physics is expected. In
particular, it is normally the strength of interaction which is said
to tune the correlation in e.g. the Hubbard model, whereas in the
Holstein model, it can be seen that both interaction strength and
phonon frequency may compete with each other to play this role.


The dynamical mean-field theory (DMFT) approach has proved successful
in treating the Holstein and other models
\cite{georges1996a,ciuchi1993a,freericks1993a}. DMFT treats the
self-energy as a momentum-independent quantity and is accurate as long
as the variation across the Brillouin zone is small. For many aspects
of the electron-phonon problem in 3D, correlations are short ranged
and DMFT can be successfully applied. The weak coupling phase diagram
was studied by Ciuchi \emph{et al.} where competing charge-order (CO)
and superconducting states were found \cite{ciuchi1993a}. Freericks
\emph{et al.} developed a quantum Monte-Carlo (QMC) algorithm
\cite{freericks1993a,freericks1994c} and examined the applicability of
several perturbation theory based techniques to the electron-phonon
problem \cite{freericks1994a,freericks1994b,freericks1998a}. The
prediction of measurable quantities away from certain well-defined
limits is severely restricted owing to difficulties inherent in the
analytic continuation. Dynamic properties such as spectral functions
can be computed in the case of static phonons \cite{millis1996a}, and
close to the static limit \cite{benedetti1998a}. Alternatively, the
limit of high phonon frequency (attractive Hubbard model) has been
studied with a QMC algorithm \cite{keller2002a}.

In the current study we are concerned with the behavior of dynamical
properties that could be measured directly with experiment. We use the
iterated perturbation theory approximation, which has been
demonstrated to be accurate for the Hubbard model, and use maximum
entropy to analytically continue the results. We compare the resulting
single-particle spectral functions over a wide range of
electron-phonon coupling strengths and phonon frequencies. The results
obtained using iterated perturbation theory (IPT) are promising and
capture generic weak and strong coupling behavior for all phonon
frequencies. At intermediate phonon frequencies, we find that
electron-phonon interactions produce a spectral function which is
simultaneously characteristic of both uncorrelated band (static) and
strongly correlated Mott/Hubbard regimes. We also find that the
competition between band-like and correlated states causes unusual
structures in the optical conductivity and resistivity. Provided a
material with high enough phonon frequency can be identified, it is
possible that such a state could be observed experimentally.

This paper is organized as follows. First, we introduce the Holstein
model, the dynamical mean-field theory and analytic continuation
techniques (section \ref{sec:formalism}). In section \ref{sec:ipt}, we
use IPT to determine the spectral functions of the Holstein model. We
compare IPT with exactly known results in the static limit. This, in
conjunction with the conclusions of Ref. \cite{freericks1994b} leads
us to argue that IPT is a reasonable approximation for the calculation
of dynamical properties in the intermediate phonon frequency
regime. We compute the density of states, optical conductivity and
resistivity, and give a heuristic explanation for their behavior.

\section{Formalism}
\label{sec:formalism}

The Holstein Hamiltonian is written as, 
%
\begin{equation}
H= -t\sum _{<ij>\sigma }c^{\dagger }_{i\sigma }c_{j\sigma } + \sum _{i\sigma }(gx_{i}-\mu )n_{i\sigma} + \sum_i\left( \frac{M\omega _{0}^{2}x_{i}^{2}}{2}+\frac{p_{i}^{2}}{2M} \right)
\label{eqn:holhamiltonian}
\end{equation}
%
The first term in this Hamiltonian represents a tight binding model
with hopping parameter $t$. The second term couples the local ion
displacement, \( x_{i} \) to the local electron density. The final
term can be identified as the non-interacting phonon Hamiltonian. \(
c^{\dagger }_{i} \)(\( c_{i} \)) create (annihilate) electrons at site
\( i \), \( p_{i} \) is the ion momentum, \( M \) the ion mass, $\mu$
the chemical potential and \( g \) the electron-phonon coupling. The
phonons are dispersionless with frequency \( \omega _{0} \).

\begin{figure}
\includegraphics[width=70mm ,height=60mm]{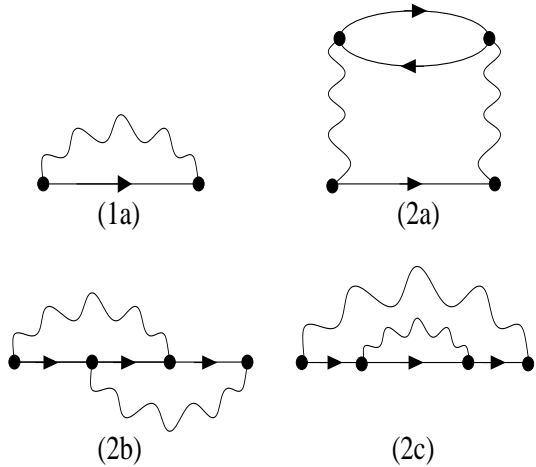}
\caption{Second order contributions to the
self-energy. Straight lines represent electron Green's functions of
the host and wavy lines phonon Green's functions.}
\label{fig:phon2o}
\end{figure}

\begin{figure}[t]
\includegraphics[width=50mm,height=100mm,angle=270]{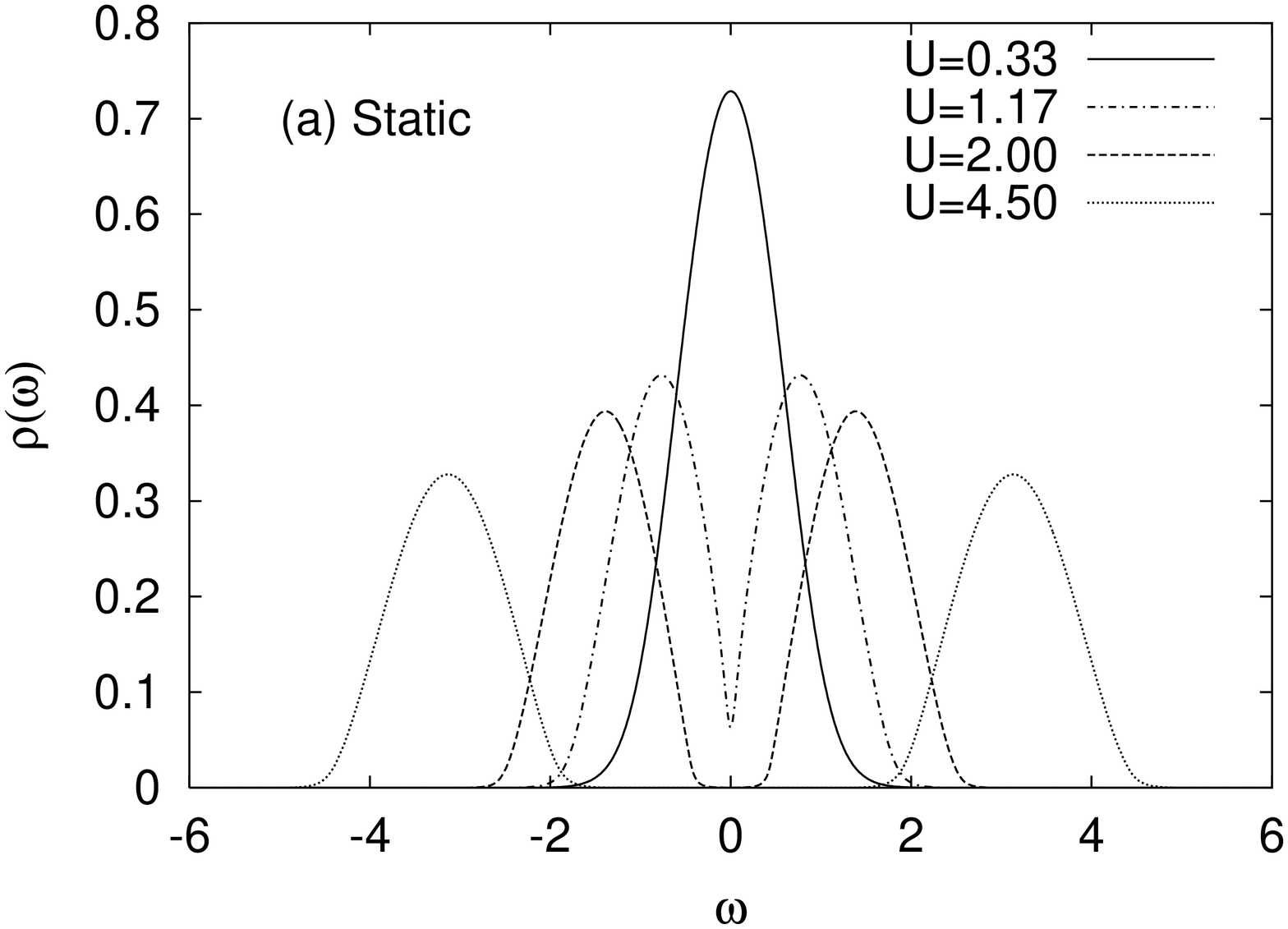}
\includegraphics[width=50mm,height=100mm,angle=270]{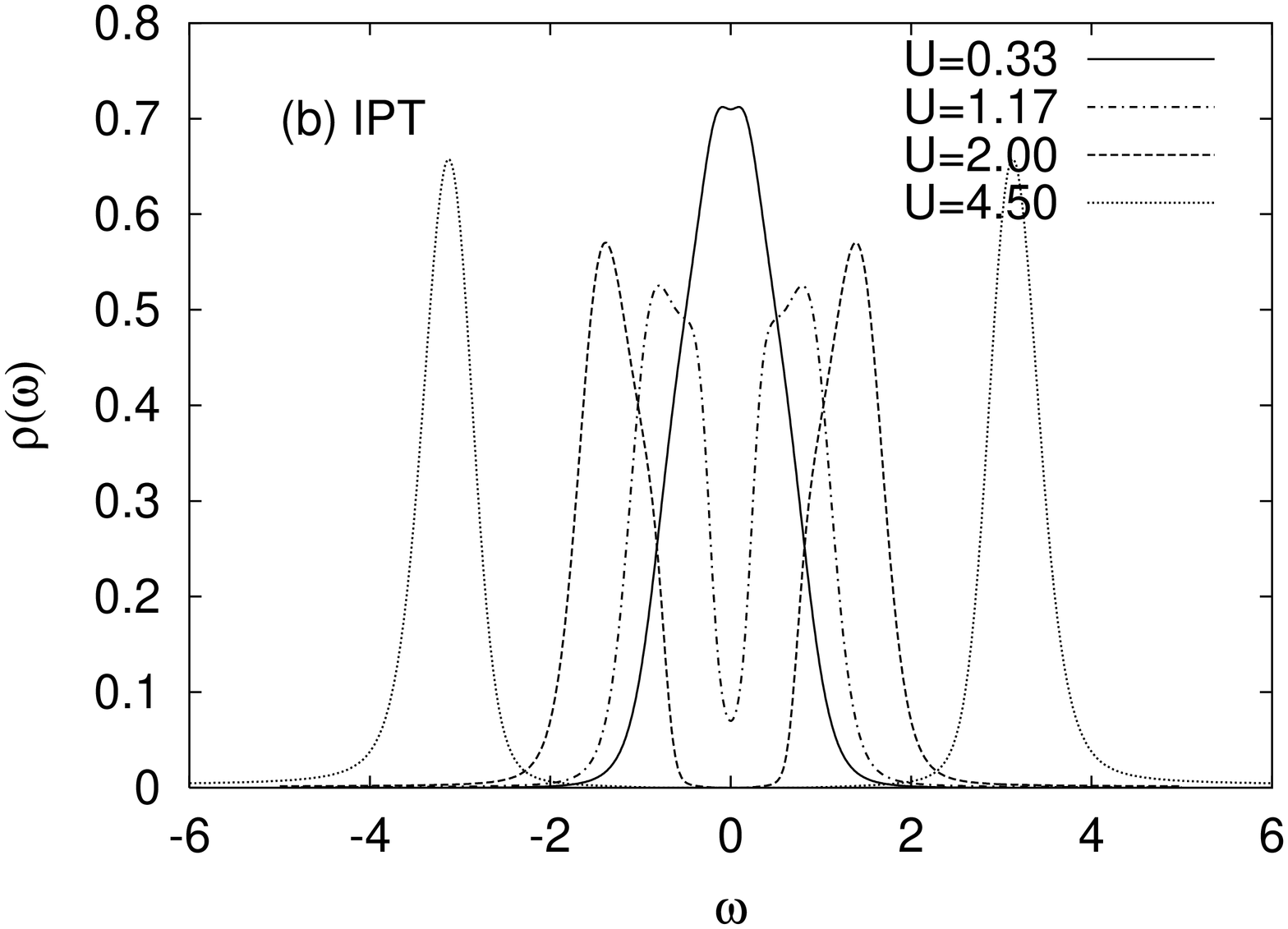}
\includegraphics[width=50mm,height=100mm,angle=270]{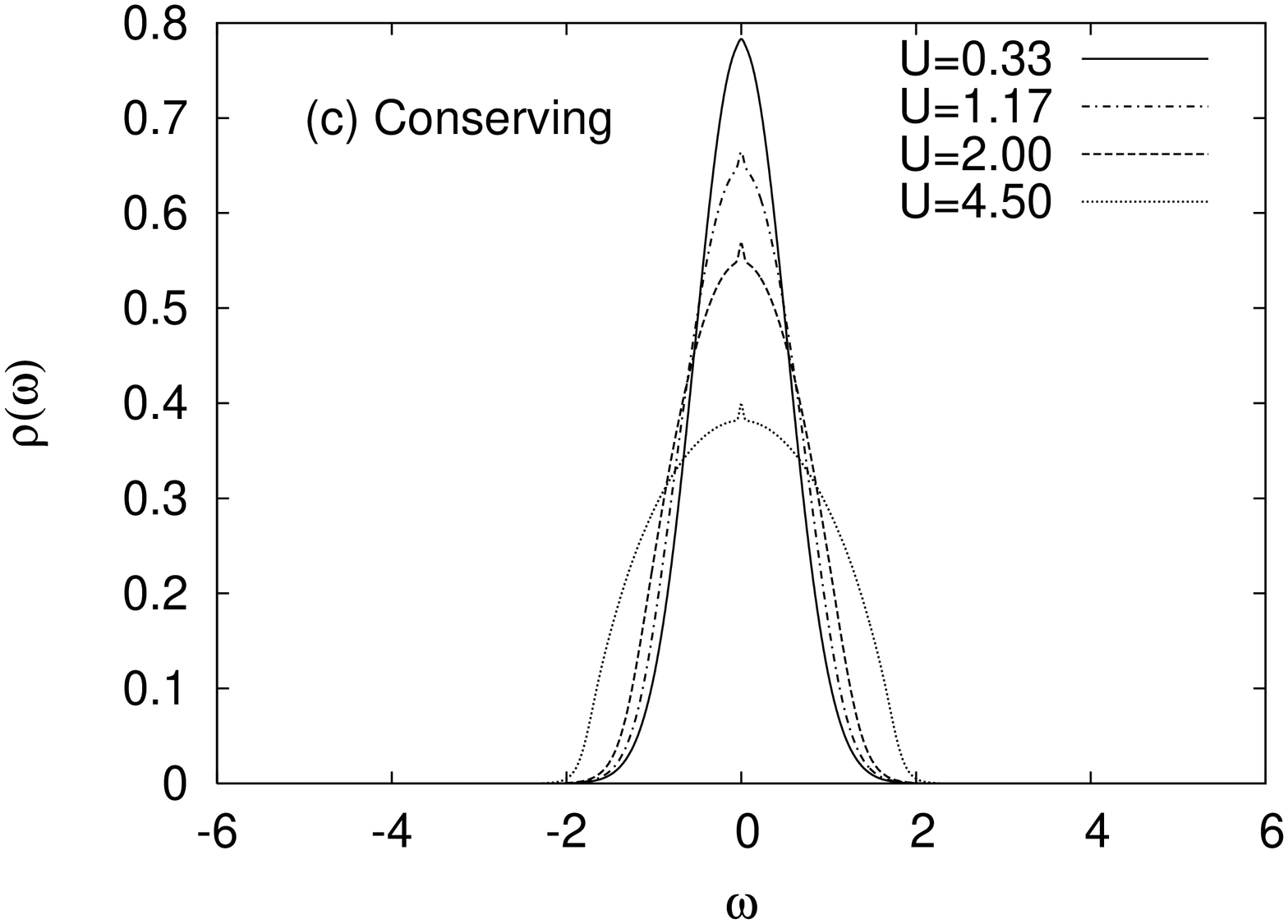}

\caption{\label{fig:cmpmillis} The spectral function in the static
limit of the half-filled Holstein model computed at temperature
$T=0.08$ (a) using the exact solution and (b) using 2nd order IPT at a
low frequency, \protect\( \omega _{0}=0.004\protect \). The IPT
solution at this small non-zero frequency is quite close to the exact
solution in the static limit. In particular, the band splitting and
the positions of the maxima agree. To contrast, panel (c) shows the
results of the approximation using the full Green's function (Diagram
2c from figure \ref{fig:phon2o} is not included to avoid
overcounting)}
\end{figure}

\begin{figure}
\includegraphics[width=50mm,height=100mm,angle=270]{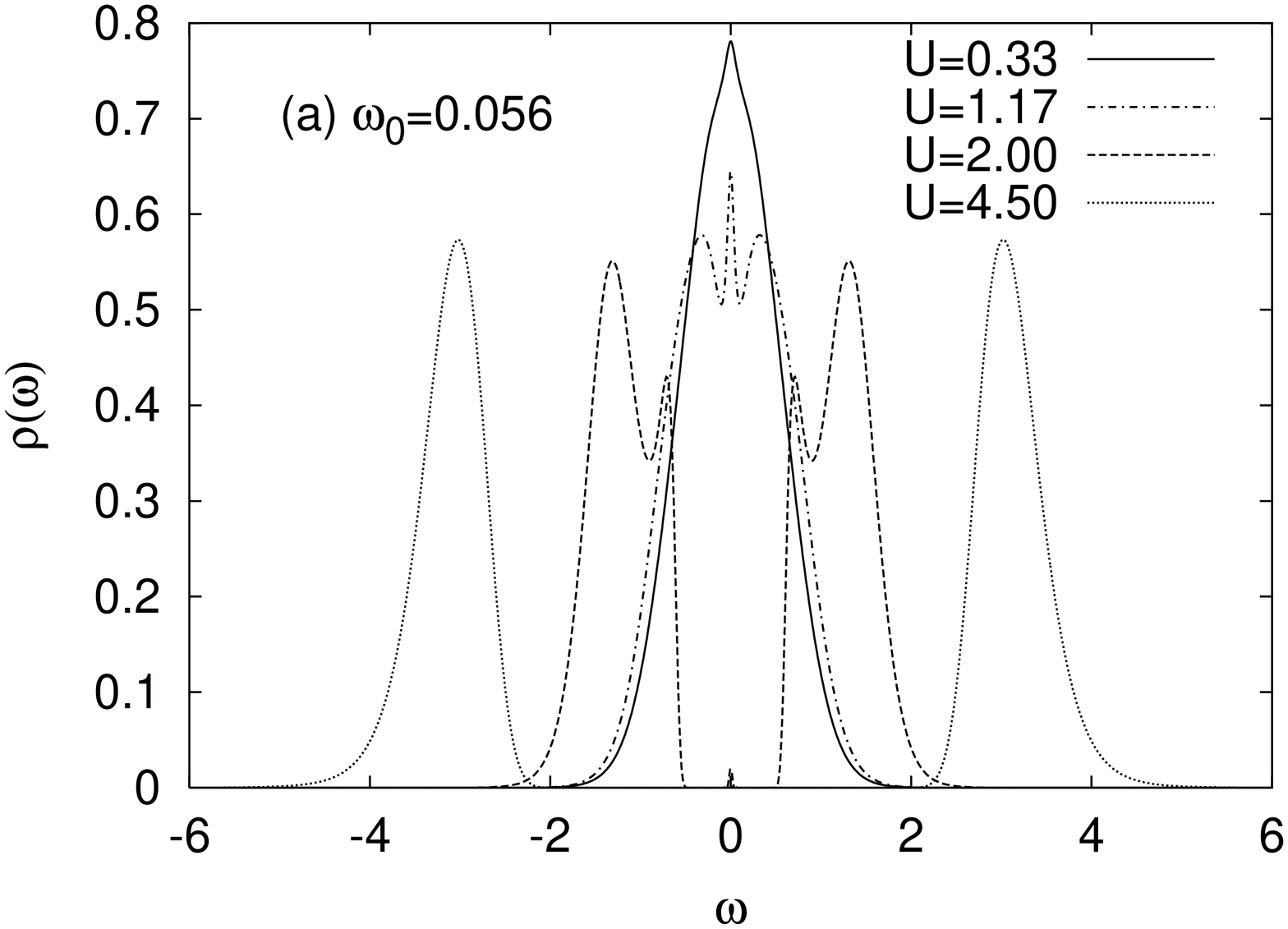}
\includegraphics[width=50mm,height=100mm,angle=270]{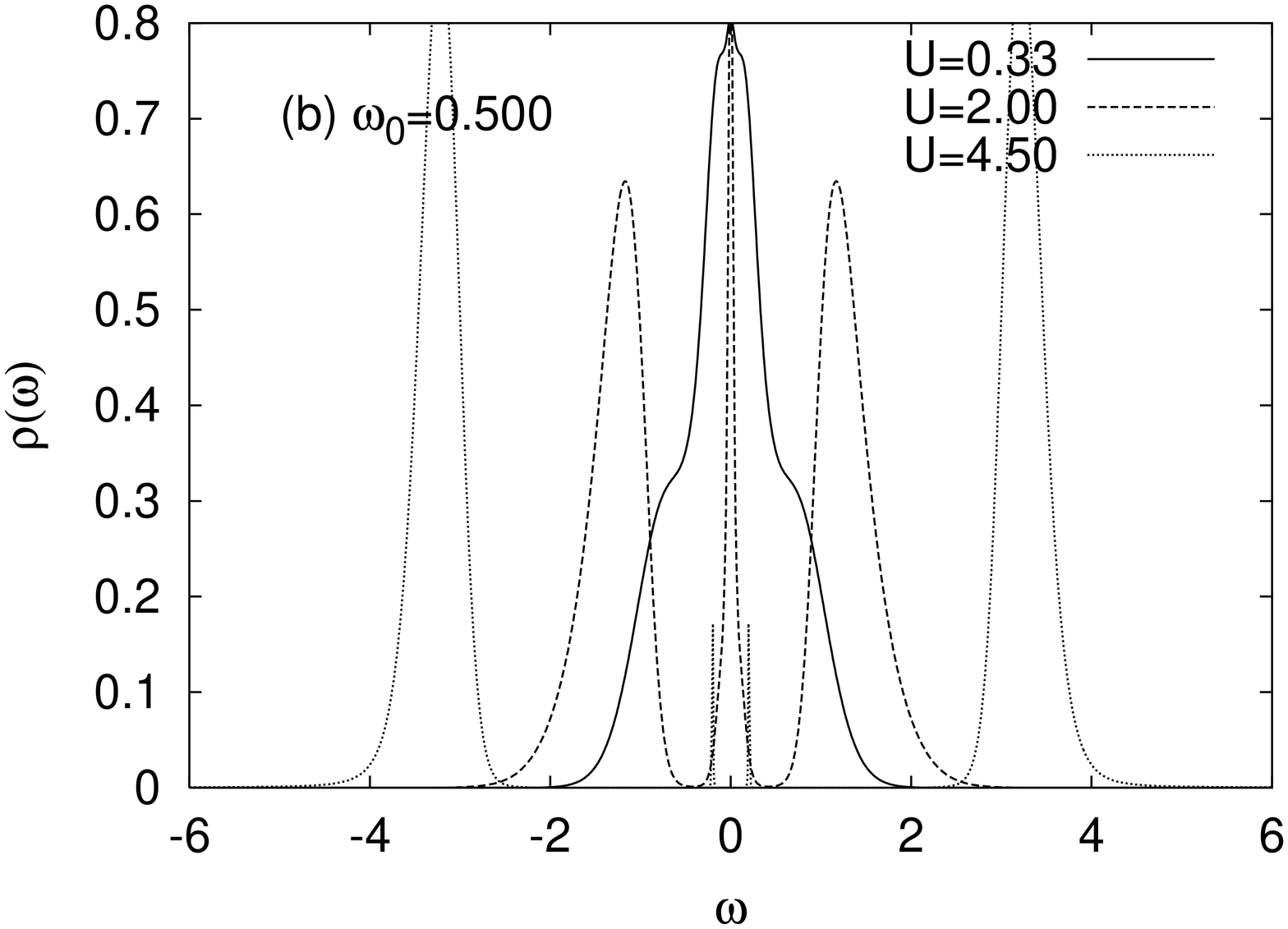}
\includegraphics[width=50mm,height=100mm,angle=270]{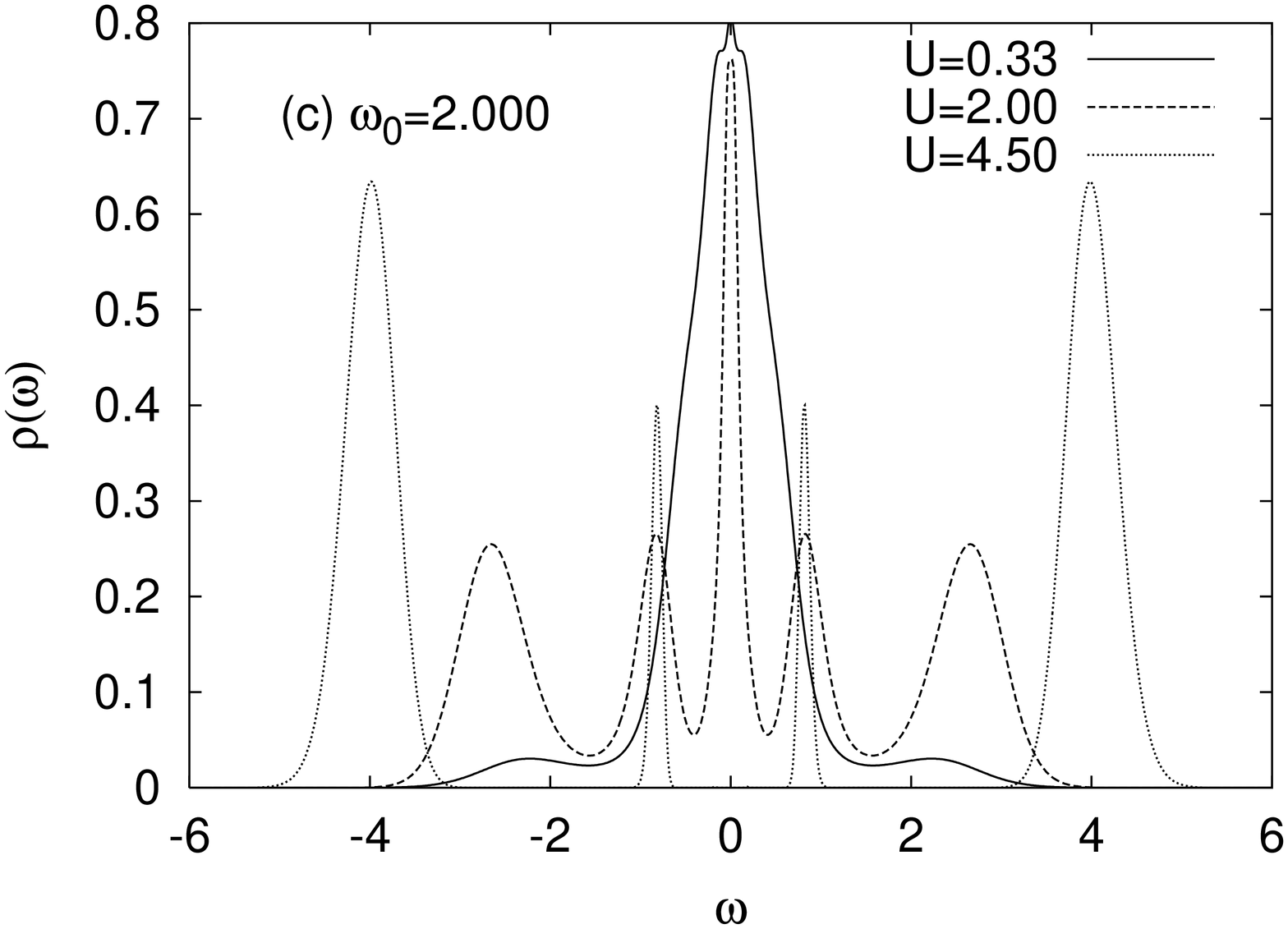}

\caption{\label{fig:ipt} Spectral functions of the half-filled
Holstein model for various electron-phonon couplings \protect\(
U\protect \), approximated using 2nd order perturbation theory at
\protect\( T=0.02\protect \) and \protect\( \omega _{0}=0.056\protect
\) (top), \protect\( \omega _{0}=0.5\protect \) (center) and
\protect\( \omega _{0}=2\protect \) (bottom). In the low frequency
limit (\protect\( \omega _{0}=0.125\protect \)), the spectral
functions are similar to those in the static limit shown in
Fig. \ref{fig:cmpmillis}, with only a small effect from the non-zero
phonon frequency. As the temperature is lower than the phonon
frequency, the central quasiparticle peak is clearly resolved for
\protect\( U\leq 2\protect \). For the intermediate frequencies
(central panel) the peak around \protect\( \omega =0\protect \) is
again clear and has a width \protect\( \sim \omega _{0}\protect \) at
low coupling.  In the gapped phase at large couplings two
band-splittings are visible.  For \protect\( \omega \gg \omega
_{0}\protect \) the band splits just as in the static limit, while for
\protect\( \omega \ll U\protect \) there is a peak at a renormalized
phonon frequency (which is less than the bare phonon frequency). In
the ungapped phases for \protect\( \omega _{0}=0.5\protect \) and
\protect\( 2\protect \), the low energy behavior is similar to that
found in the Hubbard model with a narrow quasiparticle band forming
near the Fermi energy with the value at the Fermi energy pinned to its
value in the non-interacting case.}
\end{figure}

\begin{figure}[t]
\includegraphics[width=70mm,height=100mm,angle=270]{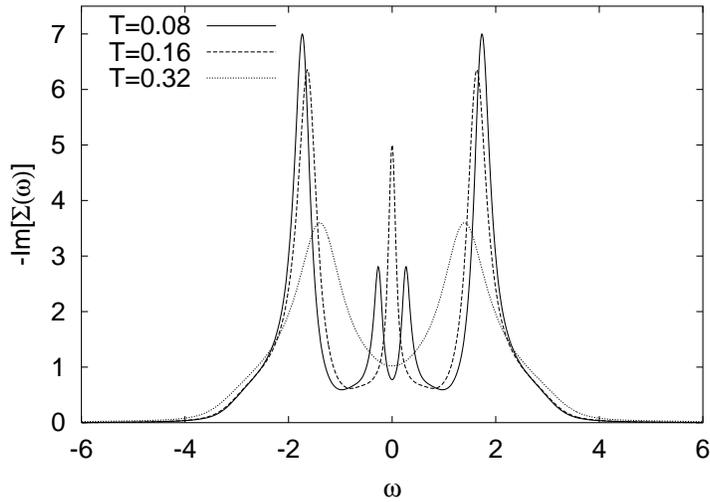}

\caption{\label{fig:se}Imaginary part of the self-energy of the half-filled Holstein
model when \protect\( U=2\protect \) and \protect\( \omega _{0}=2\protect \)
computed using IPT and analytically continued using MAXENT. At low
temperatures the low frequency behavior is Fermi-liquid like (quadratic
dependence on \protect\( \omega \protect \)) down to quite low frequencies
(at very low frequencies and low temperatures there are some inaccuracies
associated with the truncation in Matsubara frequencies). There are
peaks at the frequencies associated with the phonon energy and with
\protect\( U.\protect \) As the temperature increases the minimum
at the Fermi energy (\protect\( \omega =0\protect \)) increases as
incoherent on-site scattering in the corresponding local impurity
increases (see text). At temperatures above the characteristic
(Kondo-like) energy scale the central peak subsides and disappears.}
\end{figure}

\begin{figure}[t]
\vspace{50mm}
\includegraphics[width=70mm,height=100mm,angle=270]{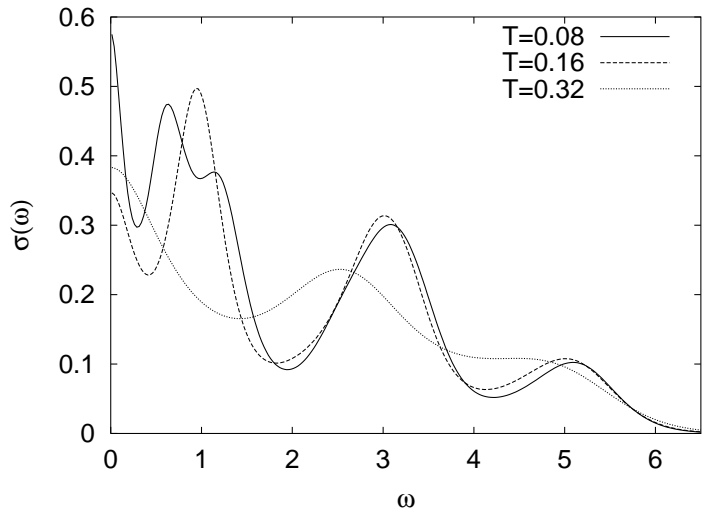}
\caption{\label{fig:oc}The real part of the optical
conductivity for a system with $U=2.0$ and $\omega_0=2.0$ for a range
of temperatures. The structure of the spectrum reflects that in the
density of states (see fig \ref{fig:ipt}. At low frequencies,
electrons may be excited within the quasiparticle resonance. The
second peak at $\omega \sim 2.0$ represents excitations from the Kondo
resonance to the large satellite (Hubbard band), and the peak at
$\omega \sim 5.0$ represents excitations between the satellites.}
\end{figure}

\begin{figure}
\includegraphics[width=70mm,height=100mm,angle=270]{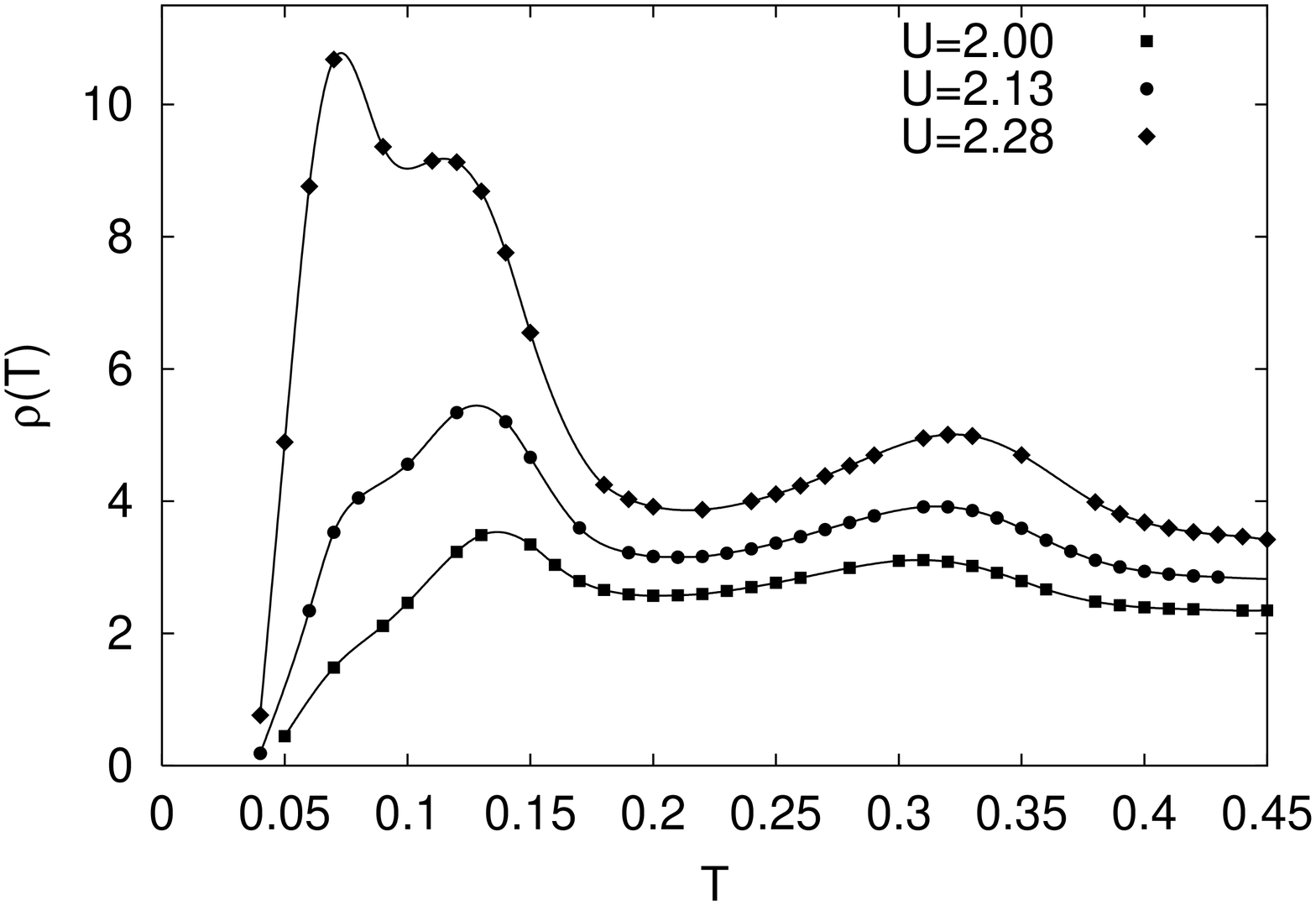}
\caption{\label{fig:res}The resistivity as a function of temperature
for the Holstein model for \protect\( \omega _{0}=2\protect \) for
varying electron-phonon coupling strengths. The resistivity is in
units of \protect\( e^{2}V/ h a^2 \protect \) with $V$ the unit cell
volume and $a$ the lattice cell spacing.  The behavior reflects what
is seen in the self-energy.  At low temperatures the behavior is
similar to that in a Kondo lattice.  The resistivity rises sharply
with temperature for temperatures smaller than the quasiparticle
bandwidth. The resistivity then drops for temperatures larger than
this lattice coherence temperature. A simple logarithmic decay with
temperature is not visible because, in addition to the Kondo-like
scattering processes, the electrons are scattered from thermally
excited phonons whose spectral weight broadens and shifts towards
lower frequencies as the temperature rises. This leads to a second
peak. In contrast, the second peak is not visible for the Hubbard
model, and indicates the presence of two energy scales in the Holstein
model.}
\end{figure}

The perturbation theory of this model may be written down in terms of
electrons interacting via phonons with the effective interaction,
\begin{equation}
\label{eqn:phononfn}
U(i\omega _{s})=-\frac{g^{2}}{M(\omega _{s}^{2}+\omega _{0}^{2})}
\end{equation}
Here, \( \omega _{s}=2\pi sT \) are the Matsubara frequencies for
bosons and \( s \) is an integer.  Taking the limit \( \omega
_{0}\rightarrow \infty \), \( g\rightarrow \infty \), while keeping
the ratio \( g/\omega _{0} \) finite, leads to an attractive Hubbard
model with a non-retarded on-site interaction \( U=-g^{2}/M\omega
_{0}^{2} \). Iterated-perturbation theory (IPT) is known to be a
reasonable approximation to the half-filled Hubbard model
\cite{georges1992b,rozenberg1992a}. Taking the opposite limit (\(
\omega _{0}\rightarrow 0 \), \( M\rightarrow \infty \), keeping \(
M\omega _{0}^{2}\equiv \kappa \) finite) the phonon kinetic energy
term vanishes, and the phonons depend on a static variable \( x_{i}
\). As such, the model may be considered as uncorrelated.


We solve the Holstein model using dynamical mean-field theory
(DMFT). DMFT freezes spatial fluctuations, leading to a theory which
is completely momentum independent, while fully including dynamical
effects of excitations. In spite of this simplification, DMFT predicts
non-trivial (correlated) physics and may be used as an approximation
to 3D models \cite{metzner1989a}. As discussed in
Ref. \onlinecite{georges1996a}, DMFT involves the solution of a set of
coupled equations which are solved self-consistently. The Green's
function for the single site problem, $G(i\omega _{n})$ can be written
in terms of the self-energy $\Sigma(i\omega_n)$ as,
\begin{equation}
\label{eqn:weissfield}
G^{-1}(i\omega _{n})={\mathcal{G}}_{0}^{-1}(i\omega _{n})-\Sigma (i\omega _{n}),
\end{equation}
 where \( \Sigma \) is a functional of \( {\mathcal{G}}_{0} \), the
Green's function for the host of a single impurity model. Here $\omega
_{n}=2\pi T(n+1/2)$ are the usual Matsubara frequencies. The
assumptions of DMFT are equivalent to taking the self-energy of the
original lattice problem to be local, hence \( G \) is also given by,
\begin{equation}
\label{eqn:greensfn}
G(i\omega _{n})=\int \frac{d\epsilon \, {\mathcal{D}}(\epsilon _{k})}{i\omega _{n}+\mu -\Sigma (i\omega _{n})-\epsilon _{k}}
\end{equation}
where \( {\mathcal{D}}(\epsilon ) \) the density of states (DOS) of
the non-interacting problem (in our case \( g=0 \)). We work with a
Gaussian DOS which corresponds to a hypercubic lattice
\cite{metzner1989a}, ${\mathcal{D}}(\epsilon)=\exp (-\epsilon
^{2}/2t^{2})/t\sqrt{2\pi }$. Equations (\ref{eqn:weissfield}) and
(\ref{eqn:greensfn}) are solved according to the following
self-consistent procedure: Compute the Green's function from equation
(\ref{eqn:greensfn}) and the host Green's function of the effective
impurity problem, ${\mathcal{G}}_0$, from equation
(\ref{eqn:weissfield}); then calculate a new self-energy from the host
or full Green's functions. In the following we will take the hopping
parameter \( t=0.5 \), which sets the energy scale.

Once the algorithm has converged, and after analytically continuing to
the real axis, response functions can be computed. We use the MAXENT
method for the determination of spectral functions from Matsubara axis
data. MAXENT treats the analytic continuation as an inverse problem
\cite{gubernatis1991a}.  The Green's function, \( G(z) \), is given by
the integral transform,
\begin{equation}
\label{eqn:hilbert}
G(z)=\int \frac{\rho (x)}{z-x}\, dx
\end{equation}
where \( \rho (x) \) is the spectral function \( (\rho (\omega
)=\mathrm{Im}[G(\omega +i\eta )]/\pi ) \).  The problem of finding \(
\rho \) is therefore one of inverting the integral transform. Since
the data for \( G_{n} \) are incomplete and noisy for any finite set
of Matsubara frequencies, the inversion of the kernel of the
discretised problem is ill-defined. The MAXENT method selects the
distribution \( \rho (x) \) which assumes the least structure
consistent with the calculated or measured data. These methods have
been extensively reviewed in the context of the inversion of the
kernel in Refs.  \onlinecite{gubernatis1991a,touchette2000a}. The
applicability to the current problem has been thoroughly tested, and
is found to be accurate.


Within the DMFT formalism, many response functions follow directly from the
one-electron spectral function and the electron self-energy (essentially
because of the neglect of all connected higher point functions apart from \(
{\mathcal{G}}_{0} \)).  Here we will be interested in the conductivity
\cite{georges1996a}:
%
\begin{equation}
\mathrm{Re}[\sigma (\omega )]=\frac{\pi^2}{\omega }\int _{-\infty }^{\infty }d\epsilon {\mathcal{D}}(\epsilon )\int _{-\infty }^{\infty }d\nu \,  \rho (\epsilon ,\nu )\rho (\epsilon ,\nu +\omega ) [f(\nu )-f(\nu +\omega )]
\label{eq:conductivity}
\end{equation}
%
 where \( f(x) \) is the Fermi-Dirac distribution. Taking the limit,
\( \omega \rightarrow 0 \), leads to the DC conductivity. (The
conductivity is in units of \( e^{2} V / h a^2 \), where $a$ is the
lattice spacing, $V$ the volume of the unit cell and $e$ and $h$ are
the electron charge and Planck's constant respectively.)

\section{Results}
\label{sec:ipt} 

In this section, we examine the validity of an approximation to the
self-energy constructed from only first and second order terms with
respect to the spectral functions calculated at very high and very low
phonon frequencies.  Finally, we calculate the optical conductivity
and resistivity.

Spectral functions are shown in figures \ref{fig:cmpmillis} and
\ref{fig:ipt}.  The perturbation theory is carried out in the host
Green's function (i.e. both electrons and phonons are bare). All
diagrams in fig. \ref{fig:phon2o} are considered,
%
\begin{equation}
\Sigma _{1\mathrm{a}}(i\omega _{n})=-UT\sum _{s}{\mathcal{G}}_{0}(i\omega _{n}-i\omega _{s})D_{0}(i\omega _{s})
\end{equation}
\begin{equation}
\Sigma _{2\mathrm{a}}(i\omega _{n})=-2U^{2}T^{2}\sum _{s,m} D_{0}^2(i\omega _{n-m})
 {\mathcal{G}}_{0}(i\omega _{m}){\mathcal{G}}_{0}(i\omega _{s}){\mathcal{G}}_{0}(i\omega _{n-m+s})
\end{equation}
\begin{equation} 
\Sigma _{2\mathrm{b}}(i\omega _{n})=U^{2}T^{2}\sum _{s,m}  D_{0}(i\omega _{m-s})D_{0}(i\omega _{n-m})
 {\mathcal{G}}_{0}(i\omega _{m}){\mathcal{G}}_{0}(i\omega _{s}){\mathcal{G}}_{0}(i\omega _{n-m+s})
\end{equation}
\begin{equation}
\Sigma _{2\mathrm{c}}(i\omega _{n})=U^{2}T^{2}\sum _{s,m} D_{0}(i\omega _{n-m})D_{0}(i\omega _{m-s})
 {\mathcal{G}}_{0}^{2}(i\omega _{m}){\mathcal{G}}_{0}(i\omega _{s})
\end{equation}
%
This also gives the correct weak coupling limit for the electronic
Green's function.

We consider first the calculation of spectral functions close to the
static and Hubbard limits. In the instantaneous limit the perturbation
theories for the Holstein and Hubbard models are equivalent. It is
well known that second order perturbation theory in the host Green's
function provides a good approximation to the Hubbard model
\cite{zhang1993a}. In the static limit, the exact solution can easily
be calculated \cite{millis1996a}. Figure \ref{fig:cmpmillis} shows
spectral functions from the exact solution, computed for a hypercubic
lattice and spectral functions, computed using 2nd order perturbation
theory at a temperature of \( T=0.08 \). The phonon frequency \(
\omega _{0}=T/20 \) was chosen so that the effects of the phonon
kinetic energy are negligible compared to thermal fluctuations. This
allows a direct comparison to be made between the exact and
approximate results. The comparison shows that the widths and
positions of the major features are closely related.

The results in Figure \ref{fig:cmpmillis} for the static limit (\(
\omega _{0}\rightarrow 0) \), together with the fact that second order
IPT is known to give reasonable results in the instantaneous limit (\(
\omega _{0}\rightarrow \infty \)), suggest that the calculation of
spectral functions should also be reliable at intermediate
frequencies. We note that Freericks \emph{et al.} also find a
reasonable agreement between the IPT and QMC self-energies at
half-filling, and that this should lead to a good agreement in the
Matsubara axis Green's function. We have therefore solved the IPT
equations for the spectral functions at intermediate frequencies. We
show the results in Fig \ref{fig:ipt}.

The results of the IPT calculations in the regime of intermediate
coupling (Fig \ref{fig:ipt}) are consistent with known results for the
limiting cases. For frequencies \( \omega \gg \omega _{0} \), the
system has the qualitative behavior of the static limit: The original
unperturbed density of states splits into two sub-bands centered around
\( \pm U/2 \). For small frequencies ($\omega \ll U, \omega
_{0}$) and interaction strength, \( U \), less than some critical
value, the system behaves as an interacting electron (Hubbard) model,
since the retarded interaction between particles \( U(i\omega _{s}) \)
(see equation \ref{eqn:phononfn}) is effectively constant for
$\omega_s\ll\omega_0$. There is then a narrow quasiparticle band at
the Fermi energy with density of states at the Fermi energy pinned at
its non-interacting value \cite{georges1996a}.  
We also note that
the results for small coupling and small frequencies are in good
agreement with those calculated using ME theory in the metallic phase
\cite{hague2001a}. 

The recent renormalization group (NRG) calculations of Meyer \emph{et
 al.} \cite{meyer2002a,meyer2002b} also report the spectral function
 in the intermediate regime. The NRG is in principle an exact method
 for solving the impurity problem onto which the DMFT equations
 map. Our results are largely consistent with theirs adding further
 support to the use of IPT in the intermediate regime.  When comparing
 with the results of Meyer \emph{et al} \cite{meyer2002a}, one should
 note that the Hamiltonian (\ref{eqn:holhamiltonian}) is exactly the
 one considered in Ref. \onlinecite{meyer2002a} but with the quantity
 $g/\sqrt{2M\omega_0} =\sqrt{U\omega_0/2}$ denoted by $g$ in
 Ref. \onlinecite{meyer2002a} and with energies measured in terms of
 the full bandwidth (instead of the half-bandwidth used here). In this
 paper, we work with the Gaussian density of states for the
 non-interacting electron DOS, whereas reference
 \onlinecite{meyer2002a} uses the semi-elliptic DOS. In general, we
 would expect the critical values for the opening of a gap to be
 larger for the Gaussian case than for the semi-elliptic case.  The
 critical coupling for the parameters in Figure \ref{fig:ipt}(c) lies
 just above $U=2.0$, corresponding in the units used in
 Ref. \onlinecite{meyer2002a} to $g=1$, compared with the critical value
 found for the semi-elliptic DOS of $g=0.69$ (note that because of the
 different energy scales $\omega_0=2$ in our results corresponds to
 $\omega_0=1$ in the units of \onlinecite{meyer2002a}). However, the shapes
 of the spectral functions are similar in both cases, with a
 five-peaked structure below and four peaked structure above the
 transition.  The peaks are narrower in the IPT results than in the
 NRG results and there is less weight in the high energy peaks.  This
 may reflect the different DOS, or inaccuracies in the NRG method at
 frequencies far from the Fermi energy resulting from the logarithmic
 discretisation, but more likely the limitations of the IPT method.

Using the method outlined in section \ref{sec:formalism} it is
possible to calculate the real-axis self-energy. The temperature
evolution of the imaginary part of the self-energy may be seen in
figure \ref{fig:se} for \( U=2 \) and \( \omega _{0}=2 \) The
self-energy at low temperatures and small frequencies shows a
quadratic (Fermi-liquid like) behavior consistent with the narrow
quasiparticle peak seen in the spectral function (Fig \ref{fig:ipt})
and develops to a broad central peak at higher temperatures. There are
also peaks corresponding to the Hubbard sub-bands. With increasing
temperature these phonon-induced peaks move together and merge into
the single central maximum associated with incoherent on-site
scattering. This peak is naturally characterized within the framework
of the self-consistent impurity model formulation of the DMFT
equations \cite{georges1996a} in terms of a Kondo resonance. In this
formulation, the dynamical mean field \( {\mathcal{G}}_{0}(\omega) \) is
written in terms of a hybridization \( \Delta (\omega ) \) between the
site orbital and a bath of conduction electrons and is therefore
equivalent to an Anderson impurity model with the added complication
that \( \Delta \) is frequency-dependent and needs to be computed
self-consistently. However, many of the properties in the metallic
state are similar to those of the Anderson impurity model. In
particular the central peak in the spectral function can be viewed as
the Kondo resonance of the impurity model.

As all connected point-functions with order higher than \(
{\mathcal{G}}_{0} \) are neglected within DMFT, the computation of
$q=0$ response functions is straightforward.  As an example we show in
fig \ref{fig:oc} the (real part of the) optical conductivity for
various temperatures with $U=2$ and $\omega_0=2$. The structure seen
in the curves reflects the structure of the density of states.  There
is a strong response at low frequencies as particle-hole pairs are
excited within the `Kondo-like' quasiparticle resonance at the Fermi
energy.  The second peak at frequencies $\omega \sim 1$ arises from
excitations between the quasiparticle resonance and the large satellite
(Hubbard band), while the third peak around $\omega \sim 5.0$ involves
excitations between the satellites.  The first dip at $\omega=0.5$ is
the signature of the small Mott bands close to the Kondo resonance and
is the feature most likely to be observable experimentally.

Also calculated is the resistivity as a function of temperature
(figure \ref{fig:res}). The curves reflect the structure of the
self-energy shown in Figure \ref{fig:se}: At low temperatures the
resistivity rises quadratically as expected for interacting
electrons. The temperature scale is given by the quasiparticle
bandwidth (`Kondo temperature'). Above this temperature the
resistivity drops as the on-site (Kondo) scattering amplitude for
electrons reduces. There is a slight second peak at higher
temperatures.  The structures in $\rho$ can be traced back to the
behavior seen in the self-energy.  This second peak is the result of
an increase in scattering off the phonons: these soften slowly with
increasing temperature and, around the second peak in the resistivity
curve, outweigh the reduction in Kondo-like scattering as the
temperature increases. This effect clearly involves a partial
cancellation between two effects and hence may be sensitive to the
accuracy of the analytic continuation, which at higher temperatures
starts from reduced information (since the majority of Matsubara
points simply show asymptotic behaviour).

\section{Summary}

We have discussed the result of changing the ratio of electron and
phonon energies as a method for tuning the amount of correlation in a
model of electron-phonon interactions. We use approximate schemes to
solve for the spectral functions of the Holstein model.  On the basis
that second-order iterated perturbation theory predicts the correct
qualitative behavior at a range of couplings in the static limit as
well as describing correctly the limit of infinite phonon frequency,
we have computed the spectral function at intermediate frequencies and
couplings. We have used an adaptation of the standard maximum entropy
scheme to obtain the spectral function, the self-energy and the
conductivity of the model by analytic continuation. These quantities
had not previously been studied.

The results for the intermediate frequency regime are consistent with
what might be expected on the basis of the limiting cases (high and
low frequencies). At energy scales smaller than \(\omega_0\), the
system shows behavior similar to that of the Hubbard model found in
the instantaneous limit \( \omega_0 \rightarrow \infty \): there is a
narrow central `Kondo-resonance' or quasiparticle band. At large
energies the model behaves as it does in the static regime with a
well-defined band splitting.  At intermediate frequencies the picture
is complicated by the interplay of the loss of coherence in the
quasiparticle band and the effective renormalization of the phonon
frequency as a function of coupling and temperature. We suggest that
if systems with anomalously large phonon frequencies and couplings
exist, then the optical conductivity should bear the hallmark of the
correlation tuned regime.

\section{Acknowledgements}

The authors would like to thank F.Essler and F.Gebhard for useful discussions.
\vspace{5mm}

\bibliographystyle{unsrt}
\bibliography{tuningholstein}

\end{document}